\documentclass{svproc}
\usepackage{graphicx}%
\usepackage{multirow}%
\usepackage{amsmath,amssymb,amsfonts}%
\usepackage{mathrsfs}%
\usepackage[title]{appendix}%
\usepackage{xcolor}%
\usepackage{textcomp}%
\usepackage{manyfoot}%
\usepackage{booktabs}%
\usepackage{algorithm}%
\usepackage{algorithmicx}%
\usepackage{algpseudocode}%
\usepackage{listings}%
\usepackage{url}
\usepackage{hyperref}

\def\orcidID#1{\unskip$^{[#1]}$} % added MR 2018-03-10

\begin{document}
\mainmatter              % start of a contribution
\title{Modelling the complex dynamics of online disinformation cascades}
\titlerunning{Modelling the complex dynamics of online disinformation cascades}  % abbreviated title (for running head)

%
\author{Guillermo Romero Moreno\orcidID{0000-0002-0316-8306} \and Valentin Robu\orcidID{0000-0002-9280-2072} \and Han La Poutr\'{e}}
\authorrunning{Guillermo Romero Moreno et al.} % abbreviated author list (for running head)

\institute{Centrum Wiskunde \& Informatica, \\ Science Park 123, 1098 XG Amsterdam, The Netherlands,\\
\email{Guillermo.Romero.Moreno@cwi.nl}}%,\\ WWW home page: \texttt{http://users.com/author1.html}}

\maketitle              % typeset the title of the contribution

\begin{abstract}
%At least 70 and at most 150 words. 

Disinformation on online social media has been on the rise in the past years.
While there is extensive literature modelling social cascades, online disinformation cascades show distinctive characteristics: they provoke debunking reactions, they spread over networks with heterogeneous connectivity, and they compete for the limited attention of users --- which recommendation algorithms may partly offset.
We propose a model including these ingredients and study it through Monte--Carlo simulations and mean--field approximations, on random graphs and an empirical Twitter network.
We find that debunking reactions are double--edged: they reduce the reach of appealing items but allow unappealing ones to reach audiences they would not otherwise reach.
Degree heterogeneity makes cascades trigger at a much lower content appeal but limits their final reach: heterogeneous networks are easy to ignite yet hard to saturate.
Limited attention cancels the effects from degree heterogeneity, although the mediation of recommendation algorithms may restore them.

\keywords{Social cascades, Social media, Rumours, Disinformation, Limited attention, Recommendation algorithms}
\end{abstract}

\section{Introduction}\label{sec1}
The recent explosion of fake news and disinformation raises major concerns about the vulnerability of the online social sphere.
This type of content is typically characterised by an emotionally arousing tone and novel or surprising claims whose sources are difficult to verify.
Such content travels fast and far through social media platforms, which provide immediate communication to a large audience: false news has been found to spread farther, faster, and more broadly than the truth \cite{Vosoughi2018}.
At the same time, the volume of content reaching each user exceeds their cognitive and attention limits. These restrictions are naturally exacerbated for highly connected users \cite{lerman_information_2016} and are mediated by the algorithms that rank what appears in each feed \cite{hodas_simple_2014}.
As the design of social media platforms is optimised for engagement, it may favour the emotionally arousing content that characterises disinformation~\cite{berger2012}, enhancing its visibility. 

Inspired by epidemiological models, early models from physics studied the spread of \textit{rumours} by adapting the \emph{Susceptible--Infected--Recovered (SIR)} model and derived analytical expressions to study how the network topology affects their diffusion \cite{moreno_dynamics_2004}. Further model refinements included competing \textit{counter--information} (or \emph{debunking}) cascades~\cite{karrer_competing_2011} or even a whole `ecology' of cascades where each rumour or piece of information needs to compete for attention \cite{weng_competition_2012}.
However, the relation between disinformation and debunking is not just competitive but also \textit{parasitic}, i.e., a debunking cascade can only emerge from places where the disinformation cascade has already reached \cite{hebert-dufresne_one_2025}.
While this type of dynamics is present in some (dis)information models \cite{Chen2011,gausen_can_2021}, a thorough understanding of the dynamics for how counter--information affects the original cascade is lacking. Furthermore, how network topology, limited attention, and recommendation algorithms jointly shape the competition between disinformation and its debunking remains unexplored.

We propose a model that mimics the dynamics of disinformation cascades in the context of online social media. 
Our model includes key features of controversial information cascades and the online social networks where they spread:
\begin{itemize}
    \item The presence of \emph{counter--information}, which can emerge with some probability from any user who receives the disinformation, while users who receive the counter--information may gain interest and believe the original disinformation item with some probability as well, creating a \textit{backfire effect}
    \item \emph{High heterogeneity in the connectivity of nodes in the network} --- online social networks such as X (formerly Twitter) exhibit long--tail distributions in the degree of their nodes, with most users having few connections and some users having a number of connections several orders of magnitude higher
    \item \emph{Limited attention}, which reduces the probability that users notice the disinformation item in inverse proportion to their number of connections.
\end{itemize}
Via Monte--Carlo simulations and mean--field approximations on Erd\H{o}s--R\'{e}nyi graphs and an empirical Twitter network, we model how these factors interact and may enable or prevent disinformation cascades for a range of model parameters.
We then examine the effect of recommendation algorithms, which may reduce attention limits by boosting the appearance of controversial information in social feeds.
Our results show that counter--information is double--edged, that degree heterogeneity makes cascades easier to trigger but harder to saturate, and that limited attention cancels these effects of heterogeneity unless recommendation algorithms restore them.

\section{Model of online disinformation cascades}
    We propose here a model of disinformation cascades that is adapted to the online social media environment.
    More particularly, we are interested in modelling when the spread of a single, new piece of disinformation is introduced into a social group, potentially by a (few) malicious actor(s) who want to influence a big part of the network.
    As the online social media environment is very fast--paced, a piece of information may burst quickly into a large cascade of re-shares, but attention may shift away as quickly, leading to an end point of the cascade.
    
    Re-sharing information in online social media incurs very little effort, and it may be enough to see a piece of information once to re-share it. 
    Likewise, once a user posts about something, they are unlikely to post about it in the near future \cite{de_domenico_anatomy_2013}.
    Therefore, our starting point is the well--known Independent Cascade model \cite{Goldenberg2001}, where active agents can activate their inactive neighbours in a single attempt with a certain probability and once all attempts are exhausted, the spreading process ends.
    The system is usually initialised with very few active agents (commonly referred to as \textit{seeds}) and the research interest is on cascade sizes depending on the network topology or seed positions \cite{Kempe2003}.
    
    We assume a population where individuals can be in one of three states:
    \begin{description}
        \item[Inactive ($I$)], i.e., individuals who may be exposed or not to the disinformation item but have not posted about it,
        \item[Spreader ($S$)], i.e., individuals who re-share / post the disinformation item, or
        \item[Denier ($D$)], i.e., individuals who publicly denounce the disinformation item.
    \end{description}
    Individuals are connected through a network that represents the connections between users in online social platforms. And
    edge between nodes $u$ and $v$ means, that when $u$ changes state from $I$ to $S$ (or $D$) and re-shares (or denounces) the disinformation item, $v$ is exposed to it and may in turn change their state to $S$ or $D$ as well.
    State changes occurs according to transition probabilities as shown in Table~\ref{tab:model}.
    The parameter $P_{SS}$ determines the probability of re-sharing the disinformation once seen, which is related to the intrinsic attractiveness or appeal of the disinformation item.
    Since the information is controversial, there is also a probability $P_{SD}\leq1-P_{SS}$ that the user denounces it publicly (i.e., transition from $I$ to $D$). Users can analogously be exposed to this counter--information and relay it, or even publicly support the original information when seeing the counter--information (i.e., backfire).
    We do not allow nodes to transition from $S$ to $D$ or vice--versa, given that it is rare that rectifications are made in public.
    For simplification of the analysis, we add a symmetric constraint to the probability of cross--transitions via a parameter~$\lambda\in[0,1]$ that we call the \textit{leakage} parameter. This parameter represents the probability that an individual would oppose the information they are receiving, i.e., transitioning to $D$ after being exposed to $S$ and vice-versa.
    \begin{table}[t]%[h]
    \centering
        \begin{tabular}{r|cc|cc}
         & change to S \, & change to D \, & \, change to S \,& change to D \\\hline
        S re-shares & $P_{SS}$ & $P_{SD}$  & $P_{SS}$ & $P_{SS} \lambda$  \\
        D re-shares & $P_{DS}$ & $P_{DD}$  & $P_{DD} \lambda$ & $P_{DD}$
        \end{tabular}
    \caption{\label{tab:model}
        Probabilities for an \textit{inactive} individual to change their state to \textit{spreader} or \textit{denier} (columns) after receiving a post from a \textit{spreader} or \textit{denier} (rows).
        On the right, a parametrisation where the \textit{leakage} $\lambda \in [0,1]$ controls cross--conversions.
    }
    \end{table}
    
    In summary, a single simulation of a cascade, representing the diffusion of a single disinformation item, is as follows:
    \begin{enumerate}
        \item The starting population is formed mostly by \textit{inactive} individuals, except for very few \textit{spreaders}.
        \item At each time step, a round of re-sharing happens: all inactive connections of a (previously unselected) random spreader or denier are selected and each of them may change state with probabilities given in Table~\ref{tab:model}.
        \item The simulation ends when all spreaders in the population have been selected once, implying that the cascade has come to an end.
    \end{enumerate}

    \subsection{Mean--field approximations of the model}
    We derive a mean--field approach that includes heterogeneity in node classes. The key idea is to track the state fractions per degree class $d$, and to account for the fact that in a heterogeneous network a high-degree node is more likely to appear as a connection of any given inactive node.
    Consider an undirected, infinite, random graph where $p_d$ denotes the degree distribution (fraction of individuals who are connected to exactly $d$ others), with mean $\langle d \rangle = \sum_d d\, p_d$. Define $I_t^{(d)}$ as the fraction of inactive individuals of degree $d$ at time $t$, $\Delta S_t^{(d)}$ as the fraction of degree-$d$ individuals newly becoming spreaders at step $t$, and $\Delta D_t^{(d)}$ as the fraction newly becoming deniers.
    
    In an uncorrelated network, a randomly chosen connection of an inactive node is not a uniformly random node --- higher--degree nodes appear as connections proportionally more often. Therefore, the probability that a random connection newly became a spreader (resp.\ denier) at step $t-1$ is the degree--weighted quantity
    \begin{equation}
        \Theta^S_{t-1} \;=\; \frac{\sum_d d\, p_d\, \Delta S_{t-1}^{(d)}}{\langle d \rangle},
        \qquad
        \Theta^D_{t-1} \;=\; \frac{\sum_d d\, p_d\, \Delta D_{t-1}^{(d)}}{\langle d \rangle}.
    \end{equation}
    Intuitively, $\Theta^S_{t-1}$ (resp.\ $\Theta^D_{t-1}$) is the probability that a uniformly chosen connection was a newly active spreader (resp.\ denier) in the previous step, weighted by the excess degree.
     
    An inactive individual of degree $d$ connects to $d$ others. Each connection that became a spreader in the previous step independently triggers a transition to $S$ with probability $P_{SS}$, and each connection that became a denier triggers the same transition with probability $P_{DS}$. The probability that none of the $d$ connections induces a transition to $S$ is therefore $(1-P_{SS})^{d\,\Theta^S_{t-1}}(1-P_{DS})^{d\,\Theta^D_{t-1}}$, and analogously for a transition to $D$.  The heterogeneous mean-field equations are then
    \begin{equation}
        \begin{array}{rcl}
            \Delta S_t^{(d)} &=& I_{t-1}^{(d)}\!\left[1
                - \left(1-P_{SS}\right)^{d\,\Theta^S_{t-1}}
                  \left(1-P_{DS}\right)^{d\,\Theta^D_{t-1}}\right] \\[8pt]
            \Delta D_t^{(d)} &=& I_{t-1}^{(d)}\!\left[1
                - \left(1-P_{SD}\right)^{d\,\Theta^S_{t-1}}
                  \left(1-P_{DD}\right)^{d\,\Theta^D_{t-1}}\right] \\[8pt]
            I_t^{(d)} &=& I_{t-1}^{(d)} - \Delta S_t^{(d)} - \Delta D_t^{(d)},\\[6pt]
            \Theta^S_t &=& \dfrac{\sum_d d\, p_d\,\Delta S_t^{(d)}}{\langle d\rangle},
            \qquad
            \Theta^D_t \;=\; \dfrac{\sum_d d\, p_d\,\Delta D_t^{(d)}}{\langle d\rangle}.
        \end{array} \label{eq:mf_nd}
    \end{equation}
    Assuming initial spreaders are seeded uniformly across degree classes, then $I_{t=0}^{(d)} = I_0, \, \Delta S_{t=0}^{(d)} = S_0 = 1-I_0,$ for all $d$, with $\Theta^S_{t=0} = S_0$ and $\Theta^D_{t=0} = 0$.
    
    Assuming small values of $P_{SS}$, in the linear (early-time) regime $I_{t-1}^{(d)}=I_0$ and $\Theta^D_{t-1}=0$, the spreading dynamics decouple and the map $\Theta^S_{t-1}\mapsto\Theta^S_t$ has the multiplicative factor
    \begin{equation}
        R_0 \;=\; P_{SS}\cdot I_0 \cdot \frac{\langle d^2\rangle}{\langle d\rangle}.
    \end{equation}

\section{Experimental Results}\label{sec2}
In this section, we aim to understand the behaviour of the model and the possibility and size of cascades under different parameter regimes.
We run experiments on two different network topologies: an Erd\H{o}s--R\'{e}nyi (ER) random graph and an empirical Twitter network.
Due to its simple topology, experiments on the ER graph serve as a baseline for understanding the spreading dynamics while removing some of the isolated effects of the network topology.
The empirical Twitter network provides a more realistic topology of online connectivity patterns, characterised by a wide heterogeneity in node connectivity --- most individuals are connected to a few others while some individuals have thousands of connections --- and other node correlations.
We use an undirected version of the Twitter network available at SNAP\footnote{\url{https://snap.stanford.edu/data/ego-Twitter.html}} \cite{mcauley_learning_2012}, which consists of a single component, $81{,}306$ nodes, and $1{,}342{,}310$ edges.
Our experiments include results from the mean--field approach --- using the average degree $\langle d\rangle$ for the ER graph and the empirical degree distribution $p_d$ from the Twitter network --- complemented by Monte--Carlo simulations of the spreading dynamics.
ER graphs were built of size $N=10^4$ and with an edge probability so that the average degree matches that of the Twitter network: $\langle d\rangle=33.02$.

\subsection{The effects of networks structure and starting fraction of seeds}
    To understand the basic behaviour of the proposed model, we start with a simple scenario where the cross--conversion parameter is $\lambda=0$, i.e., no individual seeing a post about the disinformation item will denounce it publicly and the fraction of deniers will remain at $D_t=0, \; \forall t$.
    This scenario reduces the model to the classical Independent Cascade model with a single free parameter $P_{SS}$ (or the steady--state behaviour of the SIR model).
    This baseline model allows us to understand the behaviour of the cascade dynamics depending on the network structure and the fraction of initial seeds ($S_0$).

\begin{figure}[t]
        \centering
        \includegraphics[width=0.45\linewidth]{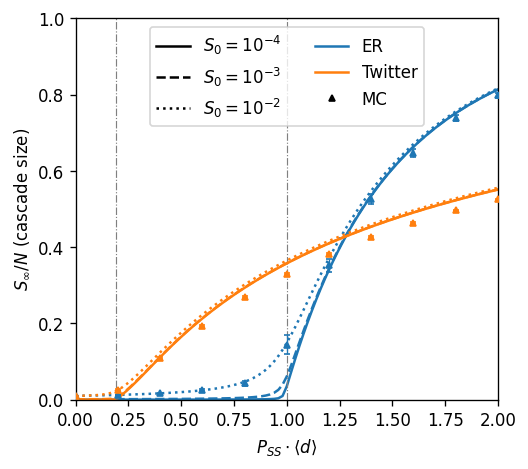}
        \includegraphics[width=0.5\linewidth]{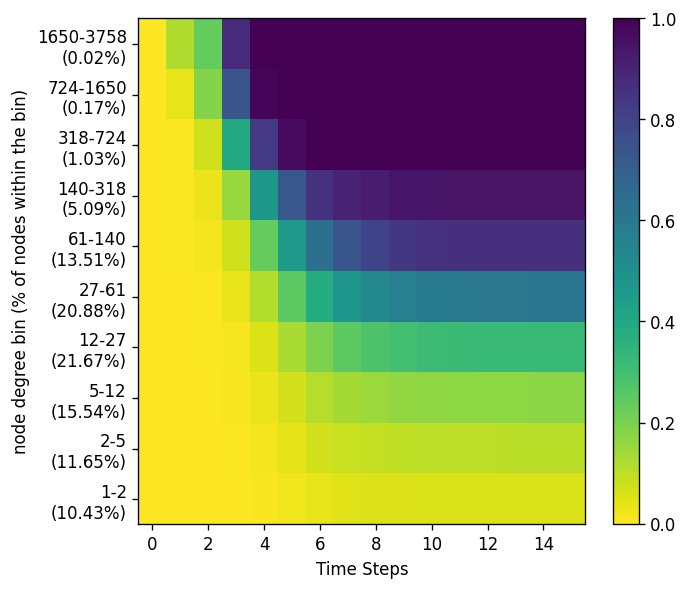}
        \caption{(a) Final cascade size ($S_{\infty}/N$) for different values of $P_{SS}$ and $S_0$, for an ER graph (blue) and an empirical Twitter network (orange), using the heterogeneous mean--field approximation from eq.~\eqref{eq:mf_nd} (lines) and Monte--Carlo simulations of the dynamics (symbols). 
        (b) Proportion of spreaders for groups of nodes in different degree bins and their evolution through time steps of a single simulation of a cascade on the Twitter network with $P_{SS}\cdot\langle d\rangle=2$ and $S_0=10^{-3}$.}
        \label{fig:no_deniers}
    \end{figure}

    We examine results of the final cascade size $S_{\infty}/N$ for different values of $P_{SS}$ and $S_0$. For experiments on the ER graph, there exists a phase transition for $P_{SS}^* \langle d\rangle\rightarrow 1$ (Fig.~\ref{fig:no_deniers}a). 
    However, this phase transition loses its sharpness when the size of the initial seed set becomes comparable to the system size (higher $S_0$).
    We also note that the effect of increasing the initial seed set is not perceivable for big or small values of $P_{SS}$ but becomes most pronounced around the transition point $P_{SS}\,\langle d\rangle\rightarrow1$.
    For this scenario, mean--field and simulation results agree very well.

    For experiments run on the empirical Twitter network, the phase transition lowers to values around $P_{SS}\langle d\rangle\approx0.19$. Indeed, if the network was infinite and its distribution was scale--free, this phase transition would be expected to approach zero and the system always be in super-criticality.
    Notably, the size of the initial seed set has a much more moderate effect in this network.
    However, even though cascades trigger at much lower values of $P_{SS}$, their size is smaller than with the homogeneous ER graph for higher values of $P_{SS}$. The reason for this difference is that, even though high--connectivity nodes are reached much quicker and able to trigger a cascade more easily, the spread cannot reach and traverse low--connectivity areas as easily (as shown in Fig~\ref{fig:no_deniers}b).
    While mean--field and simulation results match quite well, there is a consistent, small over--estimation of cascades from the mean--field results, likely due to ignoring node correlations and clustering effects that may be present in the Twitter network.

\subsection{The presence of counter--information}
    We next examine how the presence of counter--spreaders --- or deniers of the disinformation --- affects cascade sizes.
    We run experiments with fixed values of the leakage parameter at $\lambda=0.25$ (i.e., cross--conversions happen at $1/4$ of the probability of relaying the message) and a fixed initial fraction of seed spreaders at $S_0=10^{-3}$ for different values of $P_{SS}$ and $P_{DD}$, and compare them to the scenario of no deniers ($\lambda=0$) for reference.
    
\begin{figure}[t]
        \centering
        \includegraphics[width=0.9\linewidth]{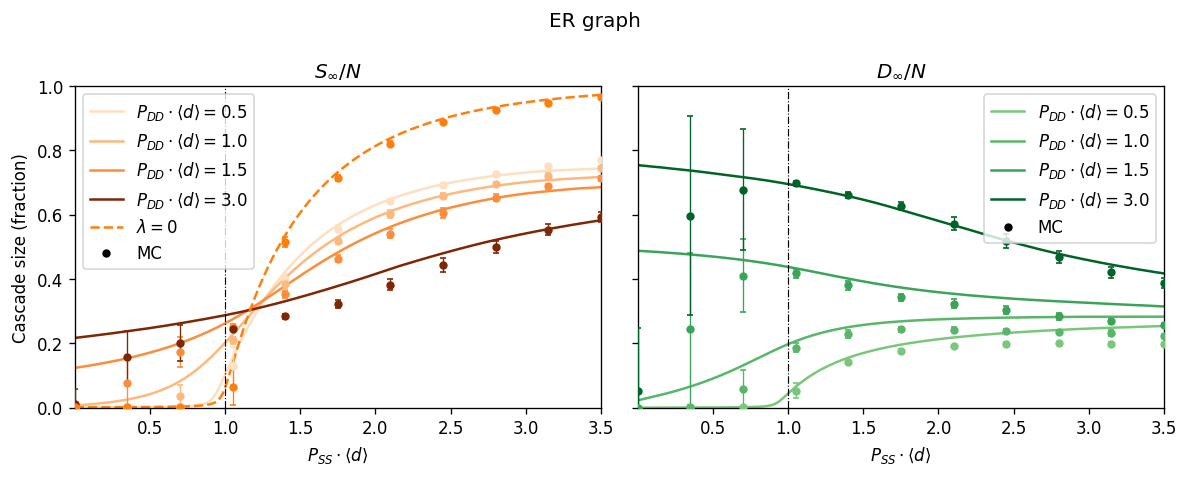}
        \includegraphics[width=0.9\linewidth]{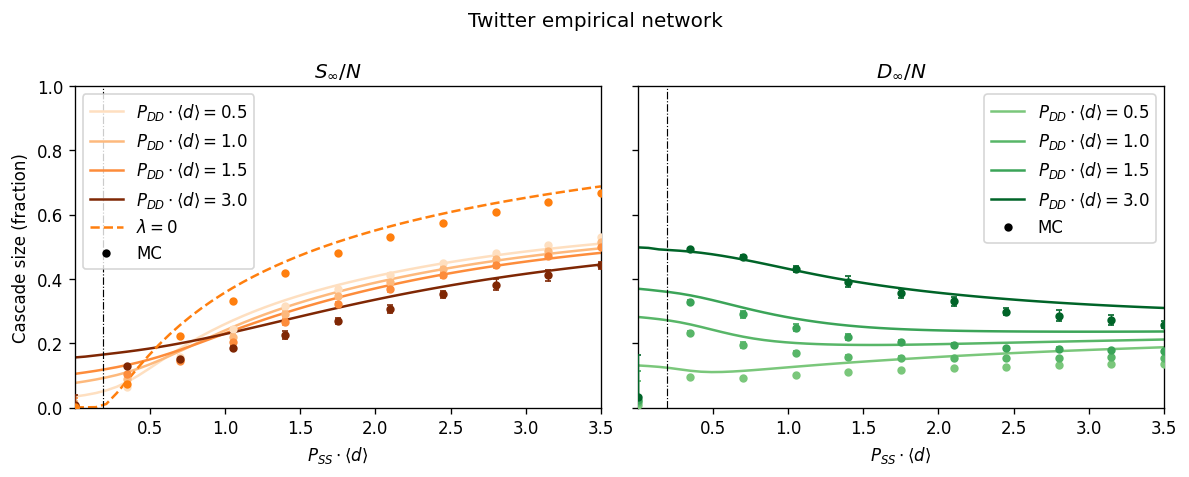}
        \caption{Final cascade size of spreaders ($S_{\infty}/N$) and deniers ($D_{\infty}/N$) for different values of $P_{SS}\cdot\langle d\rangle$ and $P_{DD}\cdot\langle d\rangle$ on an ER graph (top row) and the empirical Twitter network (bottom row), using the heterogeneous mean--field approximation from eq.~\eqref{eq:mf_nd} (lines) and Monte--Carlo simulations of the dynamics (symbols). The initial fraction of spreaders is set to $S_0=10^{-3}$ and the leakage parameter to $\lambda=0.25$. The dashed lines show scenarios with $\lambda=0$ (i.e., without deniers) as reference cases.}
        \label{fig:deniers_mf}
    \end{figure}

    We first examine the case of the ER graph (Fig.~\ref{fig:deniers_mf}, top row).
    When the spread of counter--information is in the sub-critical regime (i.e., $P_{DD}\langle d\rangle=0.5<1$), the final cascade is reduced by about 25\% compared to the case of $\lambda=0$. This reduction results from the cross-conversion due to leakage, where that lost amount directly matches with the final number of deniers ($D_{\infty}/N$), who got converted after receiving the disinformation item but whose denouncing message did not cascade.
    When spreading values are in the super-critical regime (i.e., $P_{SS}\cdot\langle d\rangle>1$), the final cascade sizes of spreaders ($S_{\infty}/N$) are further reduced as $P_{DD}\cdot\langle d\rangle$ increases, albeit modestly.
    On the contrary, an opposite effect emerges in the sub-critical regime of spreaders ($P_{SS}\cdot\langle d\rangle<1$): the final cascade of spreaders ($S_{\infty}/N$) reach fractions up to 30\% of the population while cascades would not trigger if deniers were not present (i.e., if $\lambda=0$).
    These cases result from the creation of cascades of denouncing messages reaching significant parts of the population, while they also foster cross--conversion to spreaders of the disinformation.
    Therefore, even when the presence of deniers may generally reduce the size of the cascade of spreaders, they also risk helping the message reach parts of the population where it would not have reached otherwise and find supporters there, creating a backfire effect.
    Mean--field and Monte--Carlo results agree to a lesser extent in this case, possibly due to finite--size effects. Much greater deviations can be seen in the sub-critical regime, where the distribution of cascades is bimodal: if none of the initial spreaders converts a neighbour into a denier in the initial rounds, the cascade of deniers will not trigger and the cascade of spreaders will die out.

    However, when running experiments using the more degree--heterogeneous Twitter network, the sub-critical region shrinks, reducing the possibility of a backfire effect (Fig.~\ref{fig:deniers_mf}, bottom row).
    In line with the results from Fig.~\ref{fig:no_deniers}, cascade sizes on the Twitter network are generally smaller, and less affected by the presence of deniers.

\subsection{Limited attention span and recommendations boosts}
    While a user may be connected to many other users, meaning that their posts arrive to the user's feed, the user will typically not see all of them, given that they will spend limited time using the platform and may only browse through the top posts of their feed \cite{lerman_information_2016}.
    On the other hand, the presence of contacts who are not posting about the disinformation item may also have an effect, as they also are generating content and therefore the probabilities of seeing the controversial information from the activated agents diminish.
    
\begin{figure}[!t]
        \centering
        \includegraphics[width=0.9\linewidth]{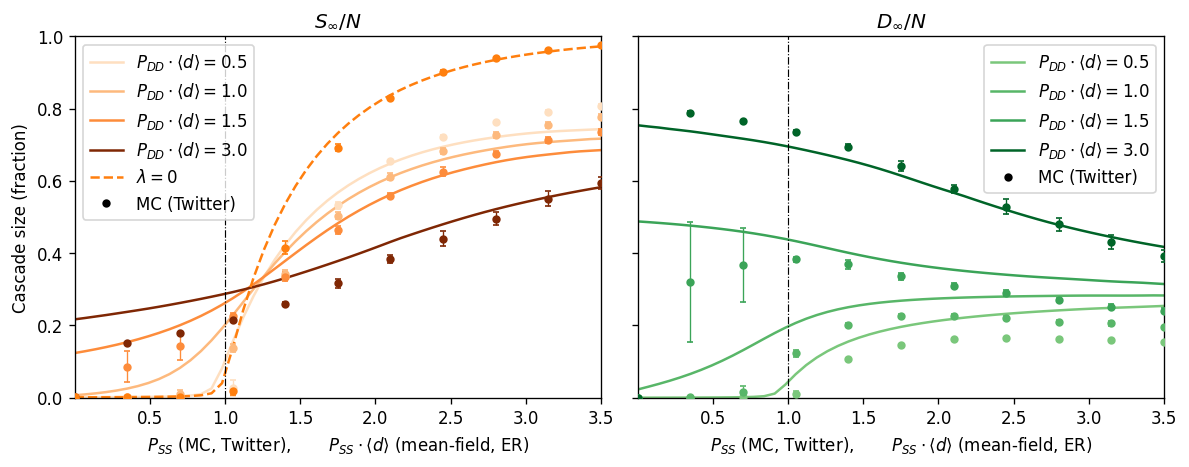}
        \includegraphics[width=0.5\linewidth]{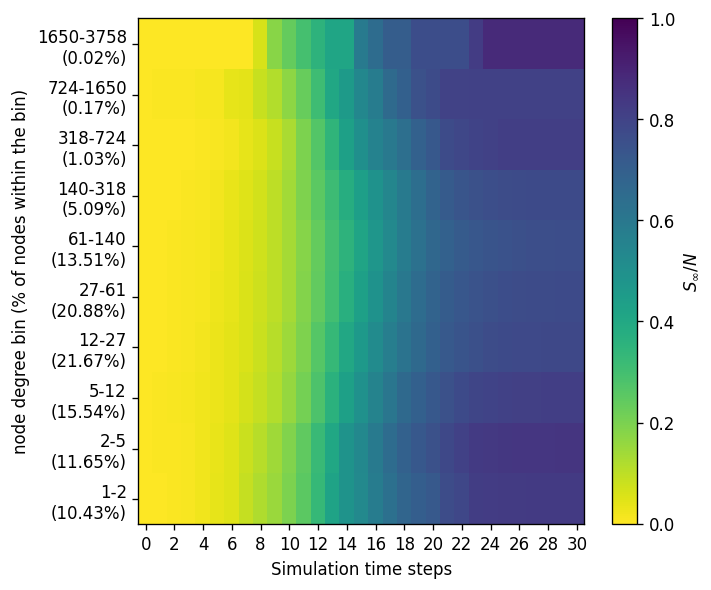}
        \includegraphics[width=0.49\linewidth]{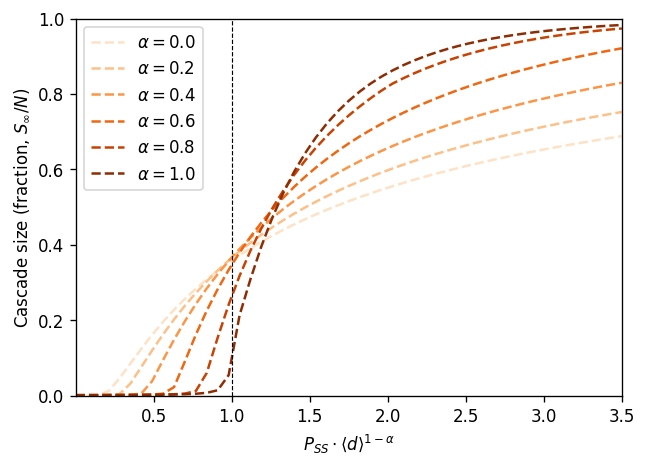}
        \caption{(a) Final cascade size of spreaders ($S_{\infty}/N$) and deniers ($D_{\infty}/N$) for different values of $P_{SS}$ and $P_{DD}$ on the empirical Twitter network under the assumption of limited attention using Monte--Carlo simulations of the dynamics (symbols) and different values of $P_{SS}\cdot\langle d\rangle$ and $P_{DD}\cdot\langle d\rangle$ on an ER graph using the mean--field approximation from eq.~\eqref{eq:mf_nd} (lines). The initial fraction of spreaders is set to $S_0=10^{-3}$ and the leakage parameter to $\lambda=0.25$. The dashed lines show scenarios with $\lambda=0$ (i.e., without deniers) as reference cases. Note that the $x$--axis uses different scales for MC and mean--field results. 
        (b) Proportion of spreaders for groups of nodes in different degree bins and their evolution through time steps of a single simulation of a cascade on the Twitter network under the assumption of limited attention, using $P_{SS}=2$ and $S_0=10^{-3}$.
        (c) Mean--field results of final cascade size of spreaders ($S_{\infty}/N$) for a range of values of $P_{SS}$ on the empirical Twitter network under the assumption of limited attention and with varying mediation from a recommendation algorithm: from no mediation ($\alpha=0$) to maximum visibility to the disinformation ($\alpha=1$). The initial fraction of spreaders is set to $S_0=10^{-3}$ and the leakage parameter to $\lambda=0$, implying no deniers are present.} 
        \label{fig:complex}
    \end{figure}

    We include these effects of limited attention by assuming that the probability of reading a post from another user is inversely proportional to the number of connections, as the other connections may be posting about other topics.
    In the model this translates to modifying the transition probabilities to make them node--dependent, by adding a multiplicative factor based on the inverse of the node degree $d_i$, i.e. $P^{(i)}=P/d_i$.
    We run additional Monte--Carlo experiments on the Twitter network with the effect of limited attention for different values of the base probabilities $P$ (Fig.~\ref{fig:complex}a). 
    In this scenario, the system recovers the phase transition at $P_{SS}\rightarrow1$ (when denier spread is in the sub-critical regime, i.e., $P_{DD}<1$), similarly to the behaviour of the model on the ER graph (Fig.~\ref{fig:deniers_mf}).
    This points to a loss of the effect of degree heterogeneity in the dynamics when limited attention is present, as the spread does not differentially reach hubs faster nor low--connectivity nodes slower, harmonising the spread through the different areas (Fig.~\ref{fig:complex}b).

    However, the presence of a recommendation algorithm may mitigate the effects of limited attention by giving preferential exposure to disinformation items, given that their novelty and emotional arousal may make them more appealing than other types of content \cite{Vosoughi2018,berger2012}.
    To emulate a recommendation algorithm that partially compensates for limited attention, we generalise the degree--scaled probabilities by setting $P^* = {P}/{d^{1-\alpha}}$,
    where $\alpha \in [0, 1]$ is an algorithmic mediation parameter where $\alpha=0$ implies no mediation and $\alpha=1$ that the disinformation item will be given maximum preferential treatment by the algorithm and always be visible to users above all other content. This is equivalent to the model from the previous sections where attention was not limited.
    Mean--field results on the empirical Twitter network using $P^*$ (Fig.~\ref{fig:complex}c) show how the level of recommendation mediation $\alpha$ interpolates between the scenarios of cancelling the effects of degree ($\alpha=0$, analogous to the scenario with limited attention) and a vanishing transition point for a heavy--tailed distribution ($\alpha=1$, analogous to results from the previous sections without attention limits).

\section{Discussion}
    Reflections on the model.
    We have proposed a model of disinformation cascades that includes three important features of controversial information in online social media: counter--information that emerges from the disinformation cascade and that may feed back into it, heavy--tailed connectivity of the social network, and limited attention, which recommendation algorithms may in turn mitigate.
    The model accommodates a different appeal of disinformation and its debunking ($P_{SS}$ versus $P_{DD}$) and the possibilities of opposing the received message (via leakage $\lambda$), while remaining analytically tractable: the heterogeneous mean--field approximation agrees well with Monte--Carlo simulations in most regimes.

    Through our simulations on synthetic Erd\H{o}s--R\'{e}nyi graphs and an empirical Twitter network, we have examined how these three features and their interaction affect the spread of disinformation stemming from a few seed accounts.
    First, the network structure shapes both whether and how far cascades spread.
    Notably, the full dynamics on the Twitter network are very similar to those from the mean--field approximation, which accounts only for its degree distribution and is otherwise random, implying that the degree heterogeneity is the main driver of these network differences.
    On the degree--heterogeneous Twitter network, cascades trigger at a much lower appeal than on the ER graph ($P_{SS}\langle d\rangle\approx0.19$ instead of $\approx1$, as predicted by the factor $\langle d^2\rangle/\langle d\rangle$ in $R_0$) and depend less on the number of initial seeds, as hubs are reached within a few steps and relay the item to large audiences --- consistent with the popularity of online content being largely driven by its largest broadcast~\cite{goel_structural_2016}.
    However, poorly connected users have fewer chances of being exposed and often remain unreached. Heterogeneous networks are thus easy to ignite but hard to saturate.

    Second, counter--information has double--edged effects on the reach of disinformation.
    When the disinformation item can cascade on its own, deniers moderately reduce the number of spreaders, mainly by diverting exposed users from relaying the item. % to denouncing it.
    When it cannot, a super-critical cascade of denouncing messages may carry it to parts of the population it would not otherwise reach, where some exposed users become spreaders.
    Debunking may thus become a vehicle for the message it opposes, as a direct consequence of the parasitic coupling between both cascades~\cite{hebert-dufresne_one_2025}.
    This backfire effect is mitigated by degree heterogeneity, which shrinks the region where the item is sub-critical, so the risk of amplifying items that would otherwise fade out is highest in more homogeneous networks.
    False news has been found to spread farther and faster than the truth~\cite{Vosoughi2018}, which makes the scenario of debunking having a higher appeal than the disinformation item to be unlikely globally. However, susceptibility to disinformation varies across the network, so locally super-critical counter--information may still arise in communities that debunk more actively.

    Third, limited attention cancels the network effects from having a heterogeneous degree distribution, much like degree--dependent weights restore the epidemic threshold of scale--free networks~\cite{karsai_nonequilibrium_2006}. 
    Limited attention thus acts as a natural brake on virality, consistent with most online information failing to spread widely \cite{lerman_information_2016,weng_competition_2012}.
    However, recommendation algorithms that preferentially surface novel and emotionally arousing content may remove this brake: increasing algorithmic mediation progressively restores the influence of hubs and lowers the point at which cascades trigger.

    Although our stylised model can help understand how some features of the social media landscape affect disinformation spread, it has yet to be validated against empirical data --- e.g., by calibration on observed distributions of cascade sizes~\cite{de_domenico_anatomy_2013,goel_structural_2016}.
    Further features of the online information environment could also be added and studied, such as heterogeneous activity rates of platform use~\cite{borge-holthoefer_emergence_2013} or individual heterogeneity in the susceptibility to believe the disinformation~\cite{tambuscio_fact-checking_2019}.
    Additionally, more sophisticated seeding strategies of disinformation could be tested, such as the targeting of high--degree nodes or simulating the attacks from coordinated inauthentic accounts and malicious bots~\cite{truong_quantifying_2024}, as well as the simulation of interventions that would limit their effect~\cite{bak-coleman_combining_2022}. 

    Our results suggest that the vulnerability of online social networks to disinformation cascades is not determined by content appeal alone, but emerges from the interplay between network heterogeneity, limited attention, algorithmic curation, and the counter--information that disinformation itself provokes.
    A better understanding of these mechanisms can inform the design of platforms and regulations that limit the potentially harmful impact of disinformation.

\bibliographystyle{spmpsci} % We choose the "plain" reference style
\bibliography{sample} % Entries are in the refs.bib file
\end{document}